\documentclass[12pt]{article}
\usepackage{graphicx}

\usepackage{fullpage}

\usepackage{amsmath,amsfonts,epsfig}
\usepackage{mathrsfs,todonotes,yfonts}  
\usepackage{calligra}
\usepackage{dsfont} 
\usepackage{stmaryrd}
\usepackage{bm}  

\usepackage{hyperref} 
 \hypersetup{
     colorlinks=true,
     linkcolor=black,  
     filecolor=back,
     citecolor=blue,      
     urlcolor=cyan,
     }
\usepackage[most]{tcolorbox}

\numberwithin{equation}{section}

\usepackage{mathabx}

\newcommand{\de}{\delta}

\newcommand{\La}{\Lambda}
\newcommand{\la}{\lambda}

\newcommand{\nn}{\nonumber}

\newcommand{\beq}{\begin{equation}}
\newcommand{\eeq}{\end{equation}}

\newcommand{\ber}{\begin{eqnarray}}
\newcommand{\eer}[1]{\label{#1}\end{eqnarray}}
\newcommand{\eero}{\end{eqnarray}}

\usepackage{accents}
\newcommand\thickbar[1]{\accentset{\rule{.7em}{.7pt}}{#1}}
\newcommand\smallthickbar[1]{\accentset{\rule{.5em}{.7pt}}{#1}}

\renewcommand{\bar}[1]{\smallthickbar{#1}}

\newcommand{\half}{{\textstyle{\frac12}}}

\newcommand{\bbx}{{\mathbb{X}}}
\newcommand{\bbxl}{{\mathbb{X}_L}}
\newcommand{\bbyl}{{\mathbb{Y}_L}}
\newcommand{\bbxlb}{{\bar{\mathbb{X}}_L}}
\newcommand{\bbylb}{{\bar{\mathbb{Y}}_L}}
\newcommand{\bbG}{{\mathbb{G}_+}}
\newcommand{\bbGb}{\bar{\mathbb{G}}_+}

\newcommand{\bbD}[1]{\mathbb{D}_{#1}}
\newcommand{\bbDB}[1]{\thickbar{\mathbb{D}}_{#1}}
\def\+{{+\!\!\!+}}

\newcommand{\re}[1] {(\ref{#1})}

\DeclareMathAlphabet{\mathcalligra}{T1}{calligra}{m}{n}
\DeclareFontShape{T1}{calligra}{m}{n}{<->s*[2.2]callig15}{}

\newcommand{\auth}

\usepackage{setspace}

\usepackage[sorting=none,
    backend=bibtex,
maxbibnames=99,
giveninits=true,
minalphanames=1,
maxalphanames=3,url=false
  ]{biblatex}
 
\bibliography{newLVM}

\renewbibmacro{in:}{}

\newbibmacro{string+doi}[1]{
  \iffieldundef{doi}{#1}{\href{http://dx.doi.org/\thefield{doi}}{#1}}}
\DeclareFieldFormat{title}{\usebibmacro{string+doi}{\mkbibemph{#1}}}
\DeclareFieldFormat[article]{title}{\usebibmacro{string+doi}{\mkbibquote{#1}}}

\begin{document}
\begin{titlepage}
	\thispagestyle{empty}
	\begin{flushright}		
	\end{flushright}
\rightline{{UUITP-07/26}}
\rightline{{YITP-SB-2026-09}}
	\vspace{35pt}	
	\begin{center}
	    {\Large\bf The Large Vector Multiplet and\\ \vspace{8pt} Gauging $(2,2)$ $\sigma$-models} 
\vspace{40pt}
		
{Dmitri Bykov$^{a,b,c,d}$\,,~~Ulf Lindstr\"om$^{e,f}$\,,~
 and Martin Ro\v cek$^{g}$}   
		
\vspace{35pt}

{\renewcommand\baselinestretch{0.5}
{\footnotesize $^{(a)}$ 
{\it Steklov
Math.~Inst.~of Russian Academy of Sciences, 119333 Moscow, Russia}}\\[0mm]
{\footnotesize $^{(b)}${\it 
Inst.~for Theoretical and Math.~Physics, Lomonosov Moscow State University, 119991 Moscow, Russia}}\\[0mm]
{\footnotesize $^{(c)}$ {\it HSE University, 6 Usacheva str., 119048 Moscow, Russia}}\\[0mm]
{\footnotesize$^{(d)}$ {\it Beijing Inst.~of Math.~Sciences and Applications (BIMSA), 
Huairou District, 101408 Beijing, China}}  }
\\[3mm]
{\footnotesize $^{(e)}$ {\it Department of Physics and Astronomy, Division of Theoretical Physics and\\[0mm] Center for Geometry and Physics, Uppsala University, Box 516, SE-75120 Uppsala, Sweden}}\\[0mm]
{\footnotesize $^{(f)}$ {\it Physics Division, National Technical University of Athens 15780 Zografou Campus, Athens, Greece}}\\[3mm]
{\footnotesize $^{(g)}$ {\it {C.N. Yang Inst.~for Theoretical Physics, Stony Brook University, Stony Brook NY 11794-3840, USA}}}
\vspace{40pt}

{ABSTRACT} 
\end{center}

The Large Vector Multiple (LVM) is the relevant gauge multiplet for gauging isometries acting on both the chiral and the twisted chiral fields in a $(2, 2)$ sigma model. Here we show that a recently proposed new gauge multiplet is a constrained or partially dualized version of the LVM. Gauging using this multiplet results in a $(2, 2)$  $\beta\gamma$ system interacting with a sigma model.

\vspace{\fill}

\noindent{\footnotesize{Email:
 bykov@mi-ras.ru, dmitri.v.bykov@gmail.com}}
 
\noindent{\footnotesize{Email: ulf.lindstrom@physics.uu.se
 }}

\noindent{\footnotesize{Email: martin.rocek@stonybrook.edu
 }}
\end{titlepage}

\section{Introduction}

Geometric quotients may often be interpreted via the introduction of gauge fields in a physics setting. The simplest example is the symplectic reduction, cf.~\cite{HoweTownsend} for a discussion in the context of mechanics. When the manifold in question admits additional structure (such as K\"ahler, hyper-K\"ahler, etc.), this is most naturally formulated in the language of supersymmetry~\cite{Lindstrom:1983rt, HKLR} (see~\cite[\S 15.4]{MirrorSymmetryBook} and \cite{LindstromReview} for a pedagogical introduction). For example, $\mathcal{N}=(2,2)$ supersymmetric sigma models in 2D involving only chiral and anti-chiral fields naturally correspond to K\"ahler target spaces~\cite{Zumino}. However, they do not exhaust all models with this amount of supersymmetry: one can as well add so-called twisted chiral~\cite{Gates} and semichiral multiplets~\cite{BuscherRocek}. From a geometric standpoint, this means entering the realm of generalized K\"ahler geometry~\cite{HitchinGeneralized, Gualtieri, GualtieriKahler} (the dictionary between the differential geometric and supersymmetric settings was conclusively set up in~\cite{LindstromOffShellComplete}; see also the reviews~\cite{LindstromReview, SevrinThompson}). Important examples of such geometries are provided by supersymmetric WZNW models~\cite{Schoutens, IvanovKimRocek, SevrinWZW}, as well as  deformations of certain integrable sigma models~\cite{Demulder, BykovLust, BKK}. In order to facilitate quotients in this generalized setup new types of vector multiplets were proposed in~\cite{LindstromVectorMultiplets} (including the non-Abelian versions in~\cite{LindstromGeneralizedNonabelian}). The multiplet relevant for the present paper is the so-called Large Vector Multiplet (LVM) that is used to gauge symmetries acting simultaneously on chiral and twisted chiral fields. 

Recently, another multiplet has been introduced~\cite{BKK}, which leads to some new interesting quotient constructions. Here we show that this gauge multiplet  has three natural interpretations: it is a constrained Large Vector Multiplet (LVM) \cite{LindstromVectorMultiplets} which can equivalently be viewed as a partial dual to the LVM (in the sense explained below) and which can further be interpreted in terms of a particular coupling of a general $\beta\gamma$ system to a sigma model \cite{LindstromBetaGamma}. We also describe the relation of the duality proposed in~\cite{BKK} to the usual LVM duality \cite{LindstromTduality}.

\section{Gauging a BiLP with the LVM}
Chiral superfields $\phi$ and twisted chiral superfields $\chi$ satisfy
\begin{align}
\bbDB{\pm}\phi=0~,~~~\bbDB{+}\chi=\bbD{-}\chi=0
\end{align}
along with their complex conjugate (cf.~\cite{LindstromReview} for details). A sigma model made from these fields has a generalized potential $K=K(\phi,\bar\phi, \chi,\bar\chi)$ and a target space geometry which is bi-Hermitian and equipped with a Local Product structure \cite{Gates}. This geometry is called a BiLP, for short. When the geometry has isometries the model can be gauged and dualized in certain cases. When
\begin{align}\label{Kahphichi}
K=K\big(\phi+\bar\phi, \chi+\bar\chi, i(\phi-\bar\phi-\chi+\bar\chi)\big)~,
\end{align}
it has an abelian shift symmetry that acts on both chiral and twisted chiral fields:
\begin{align}\label{phichiglobshift}
    \delta \phi = i\,\varepsilon\,,\quad\quad \delta \chi= i \,\varepsilon\,,\quad\quad \varepsilon\in \mathbb{R}\,.
\end{align}
It is gauged by the Large Vector Multiplet \cite{LindstromVectorMultiplets}.

Using the conventions of  \cite{LindstromTduality} the Large Vector Multiplet consists of the real superfields $V^c,V^t$ and $V'$, with transformations 
\begin{align}\label{LVM}
&\delta V^c=i(\bar\La_c-\La_c)~,~~~\delta V^t=i(\bar\La_t-\La_t)~,~~~\delta V'=\La_t+\bar\La_t-\La_c-\bar\La_c~,
\end{align}
where $\La_c,\,\La_t$ are chiral and twisted chiral superfield parameters, respectively, and 
\beq
\delta \phi=i\La_c~~,~~~~\delta \chi=i\La_t~~.
\eeq
As in \cite{LindstromTduality}, we also define the complex combinations
\begin{align}\nonumber\label{VLR}
&V_L=\half \big[-V'+i(V^c-V^t)\big]~,~~V_R=\half \big[-V'+i(V^c+V^t)\big]~,\\[1mm]
\Rightarrow~~~ &V_L=V_R-iV^t~,\qquad\qquad\qquad\,~\widebar V_L=V_R-iV^c~.
\end{align}
These transform as
\begin{align}\label{VLVRtrans}
&\delta V_L=\La_c-\La_t~,~~~\delta V_R=\La_c-\bar\La_t~.
\end{align}
The gauged model is given by 
\begin{align}\label{KG1}
K^g=K\big(\phi+\bar\phi+V^c, \chi+\bar\chi+V^t, i(\phi-\bar\phi-\chi+\bar\chi)-V'\big)~.
\end{align}
Note that $\phi-\chi=\half[(\phi+\bar\phi)-(\chi+\bar\chi)+(\phi-\bar\phi-\chi+\bar\chi)]$, and hence we may rewrite (\ref{Kahphichi}) as
\beq
\widetilde{K}\big(\phi+\bar\phi,\chi+\bar\chi,\phi-\chi,\bar\phi-\bar\chi\big):=K\big(\phi+\bar\phi, \chi+\bar\chi, i(\phi-\bar\phi-\chi+\bar\chi)\big)~,
\eeq
which leads to the equivalent form
\begin{align}\label{LVMgauged}
\widetilde{K}^g=\widetilde{K}\big(\phi+\bar\phi+V^c, \chi+\bar\chi+V^t,\phi-\chi-iV_L,\bar\phi-\bar\chi+i\bar V_L\big)~.
\end{align}
Since we use this form from here on, we drop the tilde on $\widetilde{K}$, etc.

The  spinorial field strengths for the LVM will be needed later and are given by 
\beq\label{Gs}
\bbG=\bbDB{+}V_L~,~~~~\bbGb=\bbD{+}\bar V_L~,~~~~
\mathbb{G}_-=\bbDB{-}V_R~,~~~~\bar{\mathbb{G}}_-=\bbD{-}\bar V_R~.
\eeq
It is easy to see that these are gauge-invariant w.r.t.~the transformations~(\ref{VLVRtrans}).

\section{  Gauging a BiLP with a constrained LVM}
\label{constrained}

In \cite{BKK}, gauging of the following BiLP is considered,
\begin{align}\label{DBiLP}
 K= K\big(\phi+\bar\phi, \chi+\bar\chi, \phi-\chi,\bar\phi-\bar\chi\big)~.
\end{align}
This gauging is achieved using a multiplet $(V,\bbxl )$ where $\bbxl $ is left semichiral\footnote{The idea of using semichiral fields as gauge fields first appeared in \cite{LindstromLinearizing} and was further explored in \cite{LindstromBetaGamma}.} \cite{BuscherRocek}, $\bbDB{+}\bbxl =0$, and the transformations of the gauge fields are\footnote{Here $i\La_c$ corresponds to $\Phi$ in \cite{BKK}.}
\begin{align}\label{tfs}
&\delta V^c=i(\bar\La_c-\La_c)~,~~\delta \bbxl =i(\La_c-\La_t)
\end{align}
while 
\beq\label{hatvt}
\hat V^t=V^c+\bbxl +\bbxlb~~\implies~~\de \hat V^t = i(\bar\La_t-\La_t)
\eeq 
can be used to gauge the symmetry acting on the twisted field $\chi$. The gauged Lagrangian is
  \begin{align}\label{DG}
\hat K^g= K\big(\phi+\bar\phi+V^c, \chi+\bar\chi+\hat V^t, \phi-\chi-\bbxl ,\bar\phi-\bar\chi-\bbxlb\big)~.
\end{align}
Comparing the two potentials (\ref{LVMgauged}) and (\ref{DG}), one sees that   
$iV_L$ corresponds to $\bbxl $ in gauging $\phi-\chi$ and similarly for the complex conjugate.
 This identification is also born out in comparing $\hat V^t$ to $V^t$ since
\begin{align}\label{vcvt}
\hat V^t=V^c+\bbxl +\bbxlb~ \to ~V^c+ iV_L-i\bar V_L=V^t 
\end{align} 

This can be regarded as either a partial dualization 
using the LVM gauged potential \re{LVMgauged} {\em or} a gauging with a  LVM constrained by 
\begin{equation}\label{const}
\bbG=\bbDB{+}V_L=0~.
\end{equation}
Explicitly, consider the potential \re{LVMgauged} while imposing the constraints \re{const} using Lagrange multipliers $\Upsilon_-$ and $\bar\Upsilon_-$ 
\begin{align}\label{lagM}\nn
 &K\big(\phi+\bar\phi+V^c, \chi+\bar\chi+V^t,\phi-\chi-iV_L,\bar\phi-\bar\chi+i\bar V_L\big)-i\Upsilon_-\bbG-i\bar\Upsilon_-\bbGb\\[2mm]
 &\qquad\simeq K^g-\bbyl V_L-\bbylb\bar V_L~~,~~~\bbyl :=i\bbDB{+}\Upsilon_-~.
\end{align}
Here the last line follows from partially integrating the derivatives from the definition \re{Gs}, which implies that $\bbDB+\bbyl =0$, i.e., $\bbyl$ is left semichiral. Integrating out $\bbyl$ returns the constraints \re{const} and thus the potential \re{DG}. 
Thinking of \re{lagM} as a partial duality, we integrate out $V_L$ instead. We use the relations \re{vcvt} to rewrite 
\beq\label{NB}
\widetilde{K}^g:=K\big(\phi+\bar\phi+V^c, \chi+\bar\chi+i(V_L-\bar V_L)+V^c,\phi-\chi-iV_L,\bar\phi-\bar\chi+i\bar V_L\big) =: K(A,B,C,\bar C)~,
\eeq
and integrate out $V_L$ from
\begin{align}\label{Gprim}
&\widetilde{K}^g-\bbyl V_L-\bbylb\bar V_L~.
\end{align}
We find
\beq
iK_B-iK_C=\bbyl~~,~~~-iK_B+iK_{\bar C}=\bbylb
\eeq
which gives
\begin{align}\label{Vsln}
V_L= V_L(\phi+\bar\phi+V^c,\chi,\bar\chi,\bbyl,\bbylb)~
\end{align}
under the usual convexity (concavity) conditions. The dual potential then results from substituting this $V_L$ in \re{Gprim}.

Gauge invariance guarantees that the dual potential is independent of $\chi$ (indeed, the redefinition $V_L-i\chi\to V_L$ removes the $\chi$ dependence from $\widetilde{K}^g$ as well).
The resulting Lagrangian is of the form considered in equation (6.5) of \cite{LindstromBetaGamma} (modulo swapping left and right semichiral multiplets) and therefore describes a  sigma model interacting with a $\beta\gamma$ system\footnote{Recall that a model involving only left semichiral fields and their complex conjugates describes a $\beta\gamma$ system~\cite{BuscherRocek}, see also~\cite{LindstromReview} for a review. A conventional sigma model necessarily  involves semichiral fields of both chiralities.}, gauged using the conventional real gauge superfield $V^c$.

An important difference between the LVM and its constrained version has to do with the allowed FI terms. In case of the LVM one may add to the Lagrangian the generalized FI terms
\begin{align}
    \int\,d^4\theta\,\left(\xi \,V^c+\zeta \,V_L+\widebar{\zeta}\,\widebar{V_L}\right)\,,
\end{align}
where $\xi$ is real and $\zeta$ is complex. Under the generalized gauge transformations~(\ref{LVM})-(\ref{VLVRtrans}) they lead to a shift of the Lagrangian by a total derivative.

Once one imposes the constraint~(\ref{const}), the terms proportional to $\zeta, \widebar{\zeta}$ become total derivatives themselves and therefore may be dropped. As a result, one is left with a single real FI parameter, as in the pure K\"ahler setup.

\section{  Duality}

T-duality in superspace was first discussed in \cite{Gates} for theories of chiral superfields (see also~\cite{RocekVerlinde}; in 4D, see \cite{Lindstrom:1983rt}) and extended to more general multiplets in~\cite{LindstromTduality} (see also~\cite{MerrellTduality}).

As in the  discussion above on partial dualization, full dualization of  the  potential with the LVM isometry entails adding Lagrange multiplier  terms to \re{LVMgauged} to make the gauge potentials pure gauge. In complete analogy, this results in \cite{LindstromTduality,LindstromGeneralizedNonabelian}
\begin{align}\label{dual1}
 K^g - \bbyl V_L- \bbylb \bar V_L- {\mathbb{Y}}_R V_R-\bar{\mathbb{Y}}_R \bar V_R~,
\end{align}
with both left and right semichiral fields; $\bbDB{+}\bbyl =\bbDB{-}\mathbb{Y}_R=0$. The original potential is recovered by integrating out the semichirals while the LVM dual theory results from integrating the $V$s.  

Notice that the three first terms in \re{dual1} reproduce our Lagrange multiplier form \re{lagM} of the gauged action in \cite{BKK}. Adding the two last terms (after integrating out the $\bbyl $s) is then the starting point for dualization in \cite{BKK}.  In fact, anticipating that the left semichirals set $iV_L=\bbxl $ and using  $V_R=\widebar{V}_L+iV^c$ from \re{VLR} we immediately have 
\begin{align}\label{dual2}
 K^g -i\,\mathbb{Y}_R (\bbxlb+V^c)+i\,\bar{\mathbb{Y}}_R({\bbx}_L+V^c)~.
\end{align}
which\footnote{Up to a relabeling $\mathbb{Y}_R \rightarrow i \bar Y$ if compared to the notation of~\cite{BKK}. } is the duality formula (4.10) in \cite{BKK}: 
there the dual is found by integrating out~$V^c$. 
Thus in terms of the LVM gauged potential
\re{dual1}, the $\bbyl$ Lagrange multiplier and the $V^c$  
gauge field are integrated out. This intermediate recipe corresponds to a similar use of Legendre transformations in analytical mechanics where, e.g., only a subset of the coordinates are assigned momenta (cf Routhian mechanics).

Put differently, the LVM duality, which amounts to integrating out $V_L$ and $V_R$ from~(\ref{dual1}), may be thought of as a successive implementation of two dualities\footnote{At each step one should verify that the convexity conditions on the potential are fulfilled: unlike the K\"ahler setup where ${\partial^2 \over ( \partial V^c)^2} K(\phi+\bar \phi+V^c)={\partial^2 \over \partial \phi \partial \bar{\phi}} K>0$ is satisfied automatically due to the positivity of the metric, here this is an additional constraint.}: the constrained LVM duality~(\ref{dual2}) followed by a duality on the left semichiral fields ${\bbx}_L$ that trades ${\bbx}_L$ for another semichiral field\footnote{This is a superspace version of choosing coordinate vs. momentum representation in quantum mechanics.} $\mathbb{Y}_L$~\cite{Grisaru:1997ep}. The latter is performed by imposing the semichirality constraint on $\bbx_L$ via Lagrange multipliers, as in~(\ref{Gprim}), thus reproducing~(\ref{dual1}). Upon integrating over $\bbx_L\equiv i V_L$, one obtains a dual potential depending on $\mathbb{Y}_L$.

\section{Relation to $\beta\gamma$ systems coupled to sigma models}
\label{BetaGamma}
We now return to the dual model \re{Gprim} with $V_L=V_L(\phi,\bar\phi,V^c,\bbyl,\bar\bbyl)$. As pointed out in section \ref{constrained} this does not depend on the twisted chiral fields $\chi$ and $\bbyl$ is not a gauge field, and hence can be interpreted as $\beta\gamma$ system coupled to a sigma model \cite{LindstromBetaGamma}. The same situation also arises as follows:
A field redefinition can eliminate the twisted chiral fields from the gauged action \re{DG} while at the same time making the left chiral gauge field into a gauge-invariant semichiral field $\hat{\bbx}$:
\begin{align}
\hat{\bbx}:=\bbxl +\chi-\phi~.
\end{align}
It follows from \re{hatvt} that
\beq
\hat V^t=V+\bbxl +\bbxlb=\phi+\bar\phi+V+\hat{\bbx}+\hat{\bar{{\bbx}}}~,
\eeq
and from the transformations \re{tfs} that
\beq
\delta \hat{\bbx}=0~.
\eeq
Then in these variables the potential \re{DG} is
\beq\label{DGprime}
K(\phi+\bar\phi+V^c,\hat{\bbx},\hat{\bar{{\bbx}}})
\eeq
with no twisted chiral fields and a left semichiral spectator field. As was explained in \cite{LindstromBetaGamma}, this is what results from a sigma model with potential involving a pair of chiral fields $\phi, \phi'$ and a twisted chiral fields $\chi$:\footnote{Indeed, one of the points of~\cite{LindstromBetaGamma} was that theories of semichiral superfields (such as $\beta\gamma$ systems) can arise from Kac-Moody quotients of models involving only chiral and twisted chiral fields, giving an interpretation as a kind of quotient of a {\em conventional} $\sigma$-model.}
\beq\label{KacMoodyquotientpot}
K(\phi+\bar\phi,\phi'-\chi,\bar\phi'-\bar\chi)~,
\eeq
if we gauge the shift symmetry in $\phi$ and, independently, the Kac-Moody symmetry 
\begin{align}
& \de\phi'=\la~,~~\de\chi=\la~,~~\de\bar\phi'=\bar\la~,~~
 \de\chi=\bar\la   
 \end{align}
 with
\beq
\bbDB{\pm}\la =\partial_= \la= \bbD{-}\la=
\bbD{\pm}\bar\la=\bbDB{-}\bar\la =\partial_= \bar\la=0 ~;
\eeq
for details see section 6.3 of~\cite{LindstromBetaGamma}.

\section{Summary and outlook}
The Large Vector Multiplet that gauges chiral and twisted chiral fields was introduced in~2007. As is clear from our discussion its uses have still not been fully explored. Above we discussed partial dualizing or constraining it along with field redefinitions. There are other similar constructions possible as well as applications, e.g., to quotients. Some novel applications will be given in the upcoming publication \cite{JointPaper}, and there are important issues regarding  moment maps in the context of Generalized K\"ahler Geometry (GKG) to be sorted out. The gauge multiplets for T-duality in GKG are well known \cite{Lindstrom:2007sq} as is the gauging using the LVM.  The moment maps needed for a general quotient construction are  discussed in the mathematical literature  \cite{Lin_2006},\cite{Cavalcanti:2011wu},\cite{Bursztyn_2007}. Their full implementation in superspace has not been sufficiently explored at the level of sigma models in general. It is also interesting to elucidate the role of the restricted LVM for these constructions. Apart from that, there are important issues regarding the allowed Fayet-Iliopoulos terms, presently under study.\\

As another example of the use of the partially gauged LVM, it was found in~\cite{BKK} that in certain cases, where gauging is obstructed in $\mathcal{N}=(2,2)$ superspace for theories of chiral fields alone~\cite{HullNonlinear}, the obstruction may be lifted by introducing auxiliary twisted chiral fields and moving to the generalized K\"ahler setup instead. It would be useful to develop this approach in full generality and clarify its  implications for geometry.

\vspace{5mm}
\noindent {\bf{Acknowledgements}.} The work of DB was supported by the Russian Science Foundation grant № \href{https://rscf.ru/project/7itJjWqd9pYqz47liTvCgk4qUmdtral1n3ecf41C_0I7jWkQGL8qIR8wWeq93dHAgwWeRIZfZPc~/}{25-72-10177}. DB would also like to thank A.~Kuzovchikov and S.~Kutsubin for a collaboration on related topics. UL gratefully acknowledges the hospitality of the theory group at NTUA, Athens. The work of MR was partially supported by NSF Grant \#~PHY-221053.

\vspace{0.3cm}    
    \setstretch{0.9}
    \setlength\bibitemsep{5pt}
    \printbibliography

\end{document}